\documentclass{jfm}

\ifCUPmtlplainloaded \else
  \checkfont{eurm10}
  \iffontfound
    \IfFileExists{upmath.sty}
      {\typeout{^^JFound AMS Euler Roman fonts on the system,
                   using the 'upmath' package.^^J}%
       \usepackage{upmath}}
      {\typeout{^^JFound AMS Euler Roman fonts on the system, but you
                   dont seem to have the}%
       \typeout{'upmath' package installed. JFM.cls can take advantage
                 of these fonts,^^Jif you use 'upmath' package.^^J}%
      }
  \else
  \fi
\fi

\ifCUPmtlplainloaded \else
  \checkfont{msam10}
  \iffontfound
    \IfFileExists{amssymb.sty}
      {\typeout{^^JFound AMS Symbol fonts on the system, using the
                'amssymb' package.^^J}%
       \usepackage{amssymb}%
       \let\le=\leqslant  \let\leq=\leqslant
       \let\ge=\geqslant  \let\geq=\geqslant
      }{}
  \fi
\fi

\ifCUPmtlplainloaded \else
  \IfFileExists{amsbsy.sty}
    {\typeout{^^JFound the 'amsbsy' package on the system, using it.^^J}%
     \usepackage{amsbsy}}
    {\providecommand\boldsymbol[1]{\mbox{\boldmath $##1$}}}
\fi

\newcommand\Rey{\mbox{\textit{Re}}}  % Reynolds number
\newsavebox{\astrutbox}
\sbox{\astrutbox}{\rule[-5pt]{0pt}{20pt}}

\newcommand\p{\ensuremath{\partial}}

\newcommand\etal{\mbox{\textit{et al.$\,$ }}}

\title[Effective viscosity of grease ice]
{Effective viscosity of grease ice in linearized gravity waves}

\author[G.~De~Carolis, P.~Olla and L.~Pignagnoli ]%
{G. \ns D\ls E \ns C\ls A\ls R\ls O\ls L\ls I\ls S$^1$,\ns  P. \ns
O\ls L\ls L\ls A$^2$, \break \ns L. \ns P\ls I\ls G\ls N\ls A\ls
G\ls N\ls O\ls L\ls I$^{3,4}$}

\affiliation{$^1$ ISSIA-CNR, I--70126 Bari, Italy\\[\affilskip]
$^2$ ISAC-CNR, Sez. Lecce, I--73100 Lecce, Italy \\[\affilskip]
$^3$ ISAC-CNR, I--40129 Bologna, Italy\\[\affilskip]
$^4$ Dipartimento di Matematica, Universit\'a di Milano, I--20133 Milano, Italy}

\pubyear{2003}
\volume{???}
\pagerange{1--9}
\date{?? and in revised form ??}
\usepackage{graphicx}
\usepackage{float}
\begin{document}

\def\beq{\begin{equation}}
\def\eeq{\end{equation}}
\def\d{{\rm d}}
\def\i{{\rm i}}
\def\e{{\bf e}}
\def\ex{{\rm e}}
\def\bOmega{{\boldsymbol{\Omega}}}
\def\bsigma{{\boldsymbol{\sigma}}}
\def\homega{{\hat\omega}}
\def\x{{\bf x}}
\def\bp{{\bar p}}
\def\p{{\bf p}}
\def\E{{\bf E}}
\def\A{{\cal A}}
\def\R{{\bf R}}
\def\u{{\bf u}}
\def\w{{\bf w}}
\def\U{{\bf U}}
\def\V{{\bf V}}
\def\cm{{\rm cm}}
\def\l{{\bf l}}
\def\sec{{\rm s}}
\def\Ckol{C_{Kol}}
\def\flux{\bar\epsilon}
\def\smali{{\scriptscriptstyle i}}
\def\smalfi{{\scriptscriptstyle \frac{5}{3} }}
\def\smalL{{\scriptscriptstyle{\rm L}}}
\def\smalP{{\scriptscriptstyle {\rm P}}}
\def\smalT{{\scriptscriptstyle {\rm T}}}
\def\smalE{{\scriptscriptstyle{\rm E}}}
\def\smal1n{{\scriptscriptstyle (1.1,n)}}
\def\smaln{{\scriptscriptstyle (n)}}
\def\smalA{{\scriptscriptstyle {\rm A}}}
\def\smalze{{\scriptscriptstyle (0)}}
\def\unmezz{{\scriptscriptstyle 1/2}}
\def\smaldu{{\scriptscriptstyle 2}}
\def\smaln{{\scriptscriptstyle (n)}}
\def\smalel{{\scriptscriptstyle l}}
\def\gammaP{\gamma^\smalP}
\def\shell{{\tt S}}
\def\ball{{\tt B}}
\def\nav{\bar N}
\def\micron{\mu{\rm m}}
\font\brm=cmr10 at 24truept
\font\bfm=cmbx10 at 15truept

\maketitle

\begin{abstract}
Grease ice is an agglomeration of disc-shaped ice crystals, named
frazil ice, which forms in turbulent waters of the Polar Oceans
and in rivers as well. It has been recognized that the properties of
grease ice to damp surface gravity waves could be explained in
terms of the effective viscosity of the ice slurry. This paper is
devoted to the study of the dynamics of a suspension of
disc-shaped particles in a gravity wave field. For dilute suspensions,
depending on the strength and frequency of the external wave flow,
two orientation regimes of the particles are predicted: 
a preferential orientation regime with the particles rotating in coherent
fashion with the wave field, and a random orientation regime in which
the particles oscillate around their initial orientation while 
diffusing under the effect of Brownian motion. For both motion regimes, the
effective viscosity has been derived as a function of the wave
frequency, wave amplitude and aspect ratio of the particles. Model
predictions have been compared with wave attenuation data
in frazil ice layers grown in wave tanks.

\end{abstract}

%%%%%%%%%%%%%%%%%%%%%%%%%%%%%%%%%%%%%%%%%%%%%%%%%%%%%%%%%%%%%%%%%%%%%%%%%%
\section{Introduction}
Grease ice is a thin slurry of disc-like platelets of ice
crystals, called frazil ice, which forms in supercooled waters of
the Polar Oceans under cold and windy conditions. Frazil discs
measure approximately $0.1-0.4\cm$ in diameter and $1-100\micron$
in thickness (\cite[Kivisild 1970]{kivisild}). Grease ice can
accumulate up to tens of centimeters and significantly affect
ocean surface roughness by attenuating short waves. This effect
has been widely documented by observations of early whalers.
Furthermore, synthetic aperture radar imagery of grease ice scenes
appears dark because of the suppression of the gravity-capillary
waves resonant with the incident microwave radiation ($1-10\cm$)
(\cite[Wadhams \& Holt 1991]{wadh91}).

Laboratory measurements of wave propagation in grease ice show
that wave attenuation can be explained in terms of the medium
effective viscosity (\cite[Newyear \& Martin 1997]{nm}).
\cite[Martin \& Kauffman (1981)]{mk} developed a viscous-plastic
model to explain the observed wave attenuation. They claimed that
the viscous nature of grease ice could arise from interactions
among frazil crystals leading to the presence of an energy sink in
the wave dynamics. The authors did not present any estimate of
grease ice effective viscosity from their data. On the other hand,
wave dispersion and attenuation data of \cite[Newyear \& Martin
(1997)]{nm} were consistent with a constant viscosity value,
comparable to that of glycerin at $0^\circ C$, in the range of
frequencies from $6.6$ to $9.5\sec^{-1}$ (\cite[Newyear \& Martin
1999]{nm99}). The viscosity was estimated using a two-layer wave
propagation model, which represents grease ice as a viscous fluid
superimposed on inviscid water (\cite[Keller 1998]{keller}).

As a matter of fact, the concept
of bulk viscosity for grease ice holds because the size of the
frazil particles ($\sim 0.1\cm$) is much smaller than both the
vertical scale of grease ice ($10\cm$ in the laboratory) and the horizontal
scale (from $\sim 100\cm$ in the laboratory to hundreds of meters in the
ocean) of the travelling wave (\cite[Newyear \& Martin 1999]{nm99}). 

From the theoretical point of view, it is possible to have
detailed information on the behavior of a suspension of disk-like
particles in the velocity field of a gravity wave only in the
dilute limit. In this limit, the problem becomes that of the
behavior of an individual particle in a given flow field, in the
absence of interactions with the other particles in suspension.
Iterative approaches such as that of the differential scheme
(\cite[Bruggeman 1935]{bruggeman}) can then be used to obtain
semi-quantitative informations in the high volume fraction regimes
characteristic of grease ice.

A second simplification
is obtained disregarding inertia effects at the scale of the ice platelets.
Starting from the work of Jeffery (1922), much work has been devoted to the
dynamics of an ellipsoidal particle under creeping flow conditions. The importance
of Brownian motion for the presence of an equilibrium particle 
orientation distribution,
and consequently for the existence of a uniquely defined bulk viscosity,
was already recognized in (Taylor 1923).
It turns out that, unless the particles are so small that Brownian diffusion
dominates, the external flow strongly influences the orientation distribution 
and this effect cannot be disregarded for relatively large
particles like the ice platelets. As recognized in (Bretherton 1962), this
may lead, in flow regions characterized by high strain and low vorticity,
to the possibility of fixed orientation regimes for the particles.
Only more recently, however, have time dependent situations involving ellipsoids in
suspension, 
come under scrutiny. In (Zhang \& Stone 1998), the forces and torques acting
on an oscillating disk in a quiescent fluid have been calculated. In
(Szeri \etal1992), the orientation dynamics of an ellipsoidal particle under the
effect of combined time-dependent vorticity and strain has been analyzed.

The calculation of the bulk viscosity of a dilute suspension of ellipsoids
in a  plane shear was carried on in (Leal \& Hinch 1972), in the small but
non-zero Brownian motion regime. The effective viscosity of a concentrated suspension
of aligned disks was studied in (Sundararajakumar \etal1994), using  
slender body theory arguments. In (Phan-Thien \& Pham 2000), a differential scheme
approach was used to calculate the effective viscosity in the concentrated regime
assuming random orientation of the ellipsoids. In all cases a time independent situation
was considered. 

In the present paper, we shall consider the case of gravity waves in infinitely
deep water. In this case, it is possible to pass to a reference frame
in which the flow is time independent, and this eliminates the possibility of irregular
orbits in orientation space, observed in (Szeri \etal1992)  already in the case of
simple periodic flows. 

This paper is organized as follows. In the next section, the orientation dynamics of
a Stokesian particle in a deep water gravity wave will
be elucidated. In particular, the possibility of coherent collective
motions in the suspension will be examined. In Section III, the effective viscosity
of a dilute ellipsoid suspension will be calculated, analyzing its dependence on
the wave frequency and amplitude. In section IV the merit and limitation of a creeping flow
approach to modelling the frazil ice dynamics will be discussed. In Section V, 
using a differential scheme,
the results will be extrapolated away from the dilute limit, and will be compared
with available data from wave-tank experiments. Section VI will be devoted to conclusions.

\section{Orientation of a disk-like particle in the velocity field of a deep water gravity wave}
We consider a dilute suspension of rigid oblate axisymmetric ellipsoids, supposed small enough
that inertia be negligible at the particle scale. 
We also suppose that the particles are
free of external forces or torques and that the suspension is so dilute that the effects of mutual
interaction among particles are negligible. In this dilute limit,
the rheological properties of the suspension will descend from the response of a single particle
to the time dependent wave flow.

In order to represent the wave flow, we introduce a reference frame with the origin at the water
surface, the $x_1$-axis along the direction of propagation of the wave and the $x_2$-axis
pointing vertically towards the sea-bottom. In the hypothesis of small amplitude inviscid waves
in infinitely deep water, we obtain the following velocity field:
\beq
\begin{array}{rr}
U_1 = \widetilde{U}\exp(-kx_2) \sin (kx_1-\omega t)
\\
U_2 = \widetilde{U}\exp(-kx_2) \cos (kx_1-\omega t)
\end{array}
\label{wavefield}
\eeq
where $\tilde{U}$ is a typical value for the fluid velocity in the wave field.

\begin{figure}
\begin{center}
\includegraphics[width=10cm]{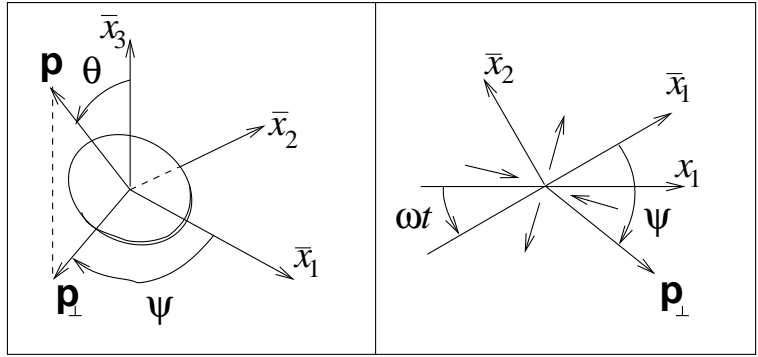}
\caption{Orientation of an ellipsoidal particle in a strain field rotating with angular velocity
$\omega$ with respect to the laboratory ($x_1$ axis).
For $0\le\homega\le 1$ the symmetry axis $\p$ is confined to the strain plane $x_1x_2$.
For $\homega\to 0$, alignment with the strain instantaneous
compressive direction $\psi=-\pi/4$ occurs.
For $0<\homega\le 1$, the symmetry axis of the particles lags
behind by a constant angle $\psi$, with $\psi=-\pi/2$ for $\homega=1$.
For $\homega>1$ no
stationary solution for $\psi$ exists, corresponding to the particle being unable to follow the
rotating strain.
}
\label{icefig1}
\end{center}
\end{figure}

The orientation dynamics in the presence of fore-aft symmetry,
will in turn be determined by the balance between the strain rate $\E$ and the vorticity
$\bOmega$ of the wave at the particle position, again provided the particle
is sufficiently small to allow linearization of the wave field on its scale.
In the case of a revolution ellipsoid, with
symmetry axis identified by the versor $\p$, the orientation dynamics will be
described by the Jeffery equation:
\beq
\dot\p = \Omega \cdot\p + G [\E\cdot\p - (\p\cdot\E\cdot\p)\p]+O((ka)^2)
\label{jeffery}
\eeq
where $G$ is the ellipsoid eccentricity defined in terms of the particle aspect ratio
$r=a/b$, where $a$ and $b$ are respectively along and perpendicular to the symmetry axis,
by means of the relation
$$
G=\frac{r^{2}-1}{r^{2}+1}
$$
For disk-like particle we have clearly $r\ll 1$ and $G\simeq -1$.
For small amplitude waves, the particle displacement will be small
with respect to $k^{-1}$ and we can approximate the instantaneous
value of the strain felt by the particle, with its value measured
at the initial position $\x$. Furthermore, for linearized waves,
the induced velocity field is confined within a region whose
thickness is of the order of the wave amplitude (defined as the
valley to crest semiheight): $\A\ll k^{-1}$. We can then approximate
$\exp(-kx_2)=1$. From Eq. (\ref{wavefield}) we
find easily the expression for the strain rate: \beq {\bf E}=k
\widetilde{U} \left(
\begin{array}{rr}
\cos (kx_1-\omega t) &    -\sin (kx_1-\omega t)  \\
-\sin(kx_1-\omega t) &    -\cos (kx_1-\omega t)  \\
\end{array}
\right)
\label{strain}
\eeq
while the vorticity $\bOmega$ is identically zero thanks to the potential flow nature
of the inviscid wave field. Equation (\ref{strain}) describes a strain field rotating
with frequency $\omega/2$ around the $x_3$ axis. Changing variables to a reference frame
rotating with the strain, the time dependence
in $\E$ disappears and a non-zero vorticity is produced:
\beq
\bar\bOmega= \frac{\omega}{2}
\left(\begin{array}{rr}
        0 &    1 \\
        -1 &   0 \\
\end{array}\right)
\label{bOmega}
\eeq
(we identify components in the rotating frame with an overbar).
%Thus, if $\E=0$, the particle would rotate with frequency $-\omega/2$ in the new reference
%frame, corresponding to a fixed orientation in the laboratory frame. 
For each value of
$x_1$, we choose the new variables in such a way that the strain rate reads:
\beq
\bar\E=k\widetilde{U}
\left(\begin{array}{rr}
       0 &     1 \\
         1 &   0 \\
\end{array}\right)
\label{strain_rot}
\eeq
corresponding to the strain expansive direction placed at $\pi/4$ with respect to the
rotating $\bar x_1$ axis. Introducing polar coordinates (see Fig. \ref{icefig1}), and
normalizing time and vorticity with the strain strength $e=k\widetilde{U}$,
Jeffery's Eq. (\ref{jeffery}) leads to the following system of equations:
\beq
\left\{ \begin{array}{l}
\dot\psi = -\homega-\cos 2\psi
 \\
\dot\theta = -\frac{1}{2}\sin 2\theta\sin 2\psi
\end{array}\right.
\label{dpsudt}
\eeq
where $\homega=-\omega/2Ge$ and $\dot f=\d f/\d\hat t$, $\hat t=-Ge t$. For $\homega<1$,
this system of equations has equilibrium
solutions $(\psi,\theta)=\frac{1}{2}(\cos^{-1}\homega-n\pi,m\pi)$. Of these, only the one
\beq
\psi=\frac{1}{2}\cos^{-1}\homega-\frac{\pi}{2}+n\pi,
\quad
\theta=\frac{\pi}{2}+n\pi
\label{equilibrium}
\eeq
is stable and is approached in a time $\sim e^{-1}$.
For $\homega>1$, instead,
choosing the time so that $\psi(0)=0$,
we have the trajectories:
\beq
\left\{ \begin{array}{l}
\tan \psi(\hat t) =-\Big(\frac{\homega+1}{\homega-1}\Big)^\frac{1}{2}\tan[(\homega^2-1)^\frac{1}{2}
\hat t],
\\
\tan\theta(\hat t)=\Big(\frac{\homega+1}{\homega+\cos 2\psi(\hat t)}\Big)^\frac{1}{2}\tan\theta(0)
\end{array}\right.
\label{trajectory}
\eeq
We thus have a high strain regime in which, as illustrated in Fig. \ref{icefig1}, the particles
in suspension are all aligned
with the local strain and rotate in coherent fashion, and a low strain regime in which
the particles do not rotate, rather, they oscillate around their natural orientation.
As it is easy to see from Eq. (\ref{trajectory}), the transition from
the low to the high strain regime is characterized by the particle spending an increasing
amount of time, as $\homega\to 1$, near $(-\pi/2,\pi/2)$. This corresponds to the rotation
period of the particle $(\homega^2-1)^{-\frac{1}{2}}$ (always measured in the rotating frame)
going to infinity,
as $\psi=-\pi/2$ becomes a fixed point for the system.
%It is interesting to note
%that the transition point $\homega=1$ corresponds to a plane shear in the rotating frame.

The linearized gravity waves theory allows us to write $e = k
\widetilde{U}$ in terms of the wave amplitude $\A$ defined as the
valley to crest semiheight, the gravitational acceleration $g$ and
the wave frequency $\omega$, starting from the dispersion
relation:
\beq
k = \frac{\omega^{2}}{g}
\label{dispersion}
\eeq
and the expression of the typical wave velocity value:
\beq
\widetilde{U}=\A \omega
\label{utilde}
\eeq
This leads to the expression for the strain strength $e$ in the
small $r$ regime corresponding to $G=-1$:
$e=\frac{\A\omega^{3}}{g}$ and to the condition for the existence
of equilibrium
\beq
\omega\ge\sqrt{\frac{g}{2 \A}}
\label{crossover}
\eeq
We thus see that in the case of
gravity waves, the aligned particle case corresponds to a high
frequency (or small wavelength) limit.
The same transition is observed experimentally in the case of 
grease ice
(see pages 307 and 308 in Martin \& Kauffman 1981), 
with the crossover frequency
at $\omega\simeq\sqrt{0.35g/\A}$.

\section{The bulk stress and the effective viscosity of the fluid-particle mixture}
The bulk stress of a dilute suspension of axisymmetric ellipsoidal
particles is given by the law (Leal \& Hinch 1972), (Hinch \& Leal 1975):
\beq 
\bsigma = 
P{\bf I} +2\mu\E+2\mu\phi \{
2A\langle\p\p\p\p\rangle:\E+2B[\langle\p\p\rangle\cdot\E+
\E\cdot\langle\p\p\rangle]+C\E+F\langle\p\p\rangle\cdot{\bf D}\}
\label{stress}
\eeq
where $\mu$
is the dynamic viscosity of the pure fluid, $P$ is the pressure,
$\phi$ is the volume fraction of the particles, $A$,$B$,$C$,$F$
are dimensionless shape coefficients and {\bf D} is a term that takes into
account Brownian motion effects. It is an open question whether
other effects, such as interaction with other particles,
could be modelled by a noise term.
The presence of this term, independently of its amplitude,
guarantees that memory of any initial
particle orientation, including unstable equilibrium points,
is lost and a statistical equilibrium state, in an $O(D^{-1})$ time,
is eventually reached.

Following (Leal \& Hinch 1972), we shall consider the small noise
limit in which $D^{-1}$ is much longer than the other timescales
of the process, which are given in dimensionless form by
$(\homega^2-1)^{-\frac{1}{2}}$. Over these timescales, the
evolution of the process will be therefore, to lowest order,
that of the unperturbed system.

The second and fourth moment of $\p$ are
calculated function of the PDF (probability density function) for
the particle orientation $\rho(\theta,\psi,t)$. The $A$,$B$,$C$ coefficients
may be obtained from (\cite[Jeffery 1922]{jeffery}), in terms of
the following elliptic integrals:
$$
\alpha'=\int_0^\infty\frac{d\lambda}{(b^{2}+\lambda)^{3}\sqrt{a^{2}+\lambda}},
\qquad
\alpha''=\int_0^\infty\frac{\lambda d\lambda}
{(b^{2}+\lambda)^{3}\sqrt{a^{2}+\lambda}}
$$
and
$$
\beta'=\int_0^\infty\frac{d\lambda}{(b^{2}+\lambda)^{2}
(a^{2}+\lambda)\sqrt{a^{2}+\lambda}},
\qquad
\beta''=\int_0^\infty\frac{\lambda d\lambda}
{(b^{2}+\lambda)^{2}(a^{2}+\lambda)\sqrt{a^{2}+\lambda}}
$$
where $a$ and $b$ identify the ellipsoid semiaxes parallel and perpendicular, respectively,
to the symmetry axis. More precisely
$$
A=\frac{\alpha''}{2b^2\alpha'\beta''}+\frac{1}{2b^{2}\alpha'}
-\frac{2}{\beta'(a^{2}+b^{2})},
\quad
B=\frac{1}{\beta'(a^{2}+b^{2})}-\frac{1}{b^{2}\alpha'}
\quad{\rm and}\quad
C=\frac{1}{b^{2}\alpha'}
$$
In the case of disk-like ($r\ll 1$) particles, disregarding $O(r)$ terms:
\beq
A=\frac{5}{3\pi r}+\frac{104}{9\pi^{2}}-1,
\quad
B=-\frac{4}{3\pi r}-\frac{64}{9\pi^{2}}+\frac{1}{2}
\quad{\rm and}\quad
C
=\frac{8}{3\pi r}+\frac{128}{9\pi^{2}}
\label{ABC}
\eeq
Notice that this value of $C$ differs at subleading order $O(1)$ from the one in 
(Leal \& Hinch 1972).
As seen in the previous section, two orientation dynamics regimes are possible and these affect
the value of the angular averages entering Eq. (\ref{stress}). We consider in detail the two
regimes below.

From the stress $\bsigma$, it is possible to calculate an effective viscosity $\bar\mu$
in terms of the viscous dissipation in the suspension, exactly as it is done with spherical
particles:
\beq
\bar\mu=\frac{1}{2}\frac{\bsigma:\E}{\E:\E}:=(1+K\phi)\mu
\label{redvisc}
\eeq
where $K$ is called the reduced viscosity for the suspension.

\subsection{Preferential orientation regimes: $0 \leq \hat{\omega}\leq 1$}

In this regime, after a relaxation time $\sim e^{-1}$,
all particles tend to align, in the rotating frame, in the direction
identified by Eq. (\ref{equilibrium}). For small diffusivities, the variance of
the distribution around these fixed points will be
$D/e$.
As already mentioned, this state of affairs corresponds, in the laboratory frame,
to the particles rotating in coherent fashion with the wave field.
The fourth and second order tensors $\langle\p\p\p\p\rangle$ and $\langle\p\p\rangle$
have a simpler form in the rotating reference frame with the $\bar x_1$ axis along $\p$.
In this new frame of reference, the rate of strain tensor $\E$ takes the following
form
$$
\bar\E =
e
\left(\begin{array}{rr}
-\sqrt{1-\hat\omega^2} &   \hat\omega  \\
\hat\omega &  \sqrt{1-\hat\omega^2}\\
\end{array}
\right)
$$
while the $\langle\p\p\p\p\rangle$ and $\langle\p\p\rangle$ tensors read:
$$
\langle\bp_i\bp_j\bp_k\bp_l\rangle = \delta_{1i}\delta_{1j}\delta_{1k}\delta_{1l}
\quad{\rm and}\quad
     \langle\bp_i\bp_j \rangle = \delta_{1i}\delta_{1j}
$$
where $\delta_{ij}$ is the Kronecker delta. Substituting into Eqs. (\ref{stress}) and
(\ref{redvisc}), the reduced viscosity coefficient $K$, is promptly obtained:
\beq
K = A(1-\hat{\omega}^{2})+2B+C
\label{Knorot}
\eeq
From Eq. (\ref{Knorot}), the dominant $O(r^{-1})$ contribution to the
viscosity is the $\homega$ dependent contribution proportional to $A$, while $2B+C=1+O(r)$.
For this reason, the reduced viscosity $K$ is characterized by
a minimum at the crossover $\homega=1$, at which $K\simeq 1$
[compare with the spherical particle value $K=5/2$, (Landau 1959)].

\subsection{Continuously rotating regime: $ \hat{\omega}\geq 1$}
In the laboratory frame this regime corresponds to the particle oscillating around its
initial orientation, while slowly diffusing with respect to angle, under the effect of
Brownian couples.
In the rotating frame, the problem can be mapped to that
of the ellipsoid in a plane shear:
the equation of motion for a particle in the rotating frame (\ref{trajectory})
is in fact identical to that of a particle with aspect ratio
\beq
s=\Big(\frac{\homega-1}{\homega+1}\Big)^\frac{1}{2}
\label{mapping}
\eeq
in a plane shear $\omega=2e$. The equilibrium distribution of an ensemble of particles
whose orientation dynamics is described by Eq. (\ref{dpsudt}), in the presence of
an isotropic Brownian couple, is then obtained from the theory of (Leal \& Hinch 1972),
whose main results are reported below.

The particle orientation is identified by the variables $\hat t$ and $c$
where $\hat t$ is defined by the first of Eq. (\ref{trajectory}) and gives the
normalized time needed, on the Jeffery orbit starting from the current values of
$\theta$ and $\psi$, to go from $\psi=0$ to the current value of $\psi$, while
$c$ obeys:
\beq
c=
\Big(\frac{\homega+\cos 2\psi}{\homega-1}\Big)^\frac{1}{2}
\tan\theta
\label{c}
\eeq
Thus, $\tan^{-1}c$ is the value of $\theta$ at $\psi=\pi/2$,
and identifies the Jeffery orbit.

In these variables, the orientation PDF can be decomposed as:
\beq
\rho(c,\hat t)=\rho(\hat t|c)\rho(c)
\label{rho}
\eeq
where, from the fact that $\hat t$ is a time along a trajectory,
$\rho(\hat t|c)\d\hat t\d c$ is also the infinitesimal solid angle element in the
variables $\hat t$ and $c$.  The marginal PDF $\rho(c)$ is given by:
\beq
\rho(c)={\rm const.}\,c[(H_4c^4+H_2c^2+H_0)F]^{-\frac{3}{4}}
\label{rho(c)}
\eeq
where
$$
H_4=s^2+1,
\quad
H_2=\frac{1}{4}s^2+\frac{7}{2}+\frac{1}{4s^2}
\quad
H_0=\frac{1}{s^2}(s^2+1)
$$
and
\beq
F=
\left\{\begin{array}{l}
\Big[\frac{2H_4c^2+H_2-S}{2H_4c^2+H_2+S}\Big]^\frac{4-H_2}{S}
\qquad\qquad\qquad\qquad\qquad\qquad\
H_2^2>4H_4H_0,
\\
\exp\Big[\frac{2(H_2-4)}{2H_4c^2+H_2}\Big]
\qquad\qquad\qquad\qquad\qquad\qquad\qquad
H_2^2=4H_4H_0,
\\
\exp[2S^{-1}(H_2-4)\tan^{-1}S^{-1}(2H_4c^2+H_2)]
\qquad
\nonumber
H_2^2<4H_4H_0,
\end{array}\right.
\label{cases}
\eeq
where $S=|H_2^2-4H_4H_0|^\frac{1}{2}$.

\begin{figure}
\begin{center}
\includegraphics[width=10cm]{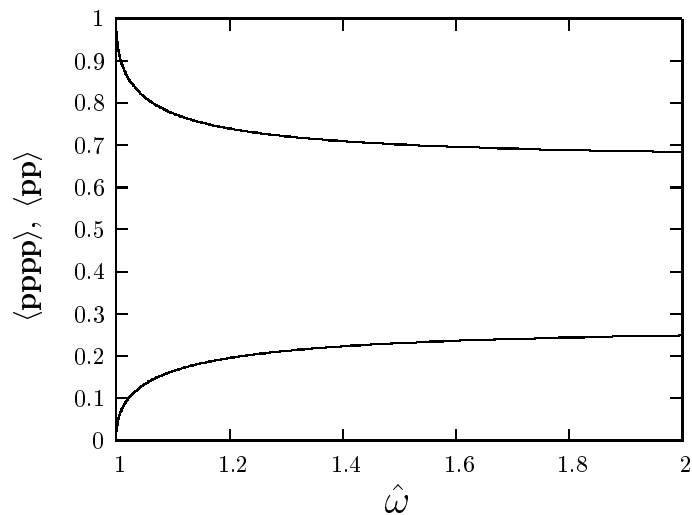}
\caption{Plot of $\langle\sin^4\theta\sin^22\psi\rangle$ (bottom) and
$\langle\sin^2\theta\rangle$ (top) vs. $\homega$ in the low strain range.
}
\label{icefig2}
\end{center}
\end{figure}

\begin{figure}
\begin{center}
\includegraphics[width=10cm]{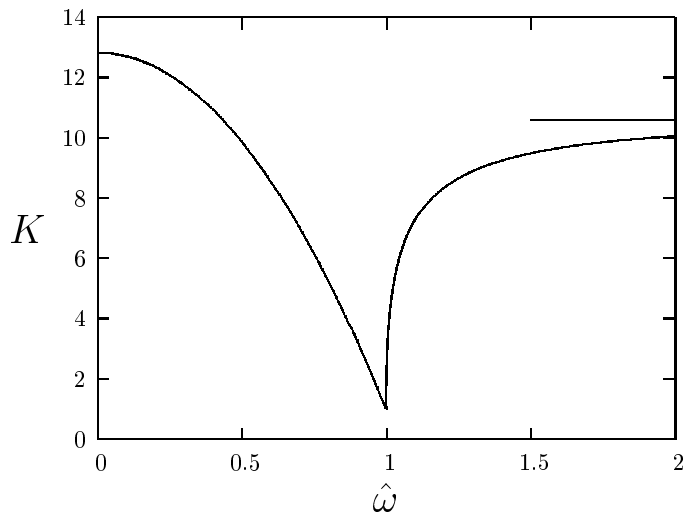}
\caption{
Reduced viscosity vs.  normalized frequency $\homega\simeq\omega/2e$ for a disk-like particle
with aspect ratio $r=0.045$. From $\homega\propto\omega^{-2}$, the long wave regime
corresponding to random particle orientation, occurs for $\homega>1$ and the short wave
one, corresponding to coherent motion, for $\homega<1$. The horizontal line to the right
gives the asymptotic value $K(\homega\to\infty)$.
}
\label{icefig3}
\end{center}
\end{figure}

The PDF $\rho(c)$ is actually the only thing that we need, since the averages along
the orbits of the tensors $\p\p$ and $\p\p\p\p$ are already available (Jeffery 1922).
In fact, from Eq. (\ref{redvisc}), the reduced viscosity can be written as
\beq
K=A\langle\sin^4\theta\sin^22\psi\rangle+2B\langle\sin^2\theta\rangle+C
\label{Krot}
\eeq
but, from (Jeffery 1922):
$$
\langle\sin^4\theta\sin^22\psi|c\rangle=\frac{2s^2}{(s^2-1)^2}\Big[
\frac{c^2(s^2+1)+2}{[(c^2s^2+1)(c^2+1)]^\frac{1}{2}}-2\Big]
$$
and
$$
\langle\sin^2\theta|c\rangle=1-\frac{1}{[(c^2s^2+1)(c^2+1)]^\frac{1}{2}}
$$
Completing the averages by means of Eq. (\ref{rho(c)}), leads to behaviors for
$\langle\sin^4\theta\sin^22\psi\rangle$ and $\langle\sin^2\theta\rangle$ shown
in Fig. \ref{icefig2}.
Substituting into Eq. (\ref{Krot}) with the expressions for the
coefficients $A$, $B$ and $C$ provided by Eq. (\ref{ABC}), allows to determine
the reduced viscosity of a dilute suspension of ellipsoidal particles, for arbitrary
values of the aspect ratio $r$ and of the reduced frequency $\homega$.
As in the continuously rotating regime, we see that the effective viscosity grows away
from the crossover at $\homega=1$, with the dip becoming more pronounced as the aspect
ratio $r$ is sent to zero.
This is illustrated in Fig. \ref{icefig3}, in the case of a disk-like particle with a value
of the aspect ratio in the range characteristic for frazil ice.
Notice that the asymptotic regime of
random particle orientation $\rho(\theta,\psi)=\frac{\sin\theta}{4\pi}$, leading to the
expression for the reduced viscosity (\cite[Phan-Thien \& Pham 2000]{pt})
$$
K=\frac{4}{15}A+\frac{4}{3}B+C
$$
is obtained already for relatively small values of the reduced frequency $\homega\simeq 2$.

We remark that, using the expression for $C$
given in (Leal \& Hinch 1972), would have produced an unphysical negative value of the
reduced viscosity $K$ at the crossover.

\section{Grease ice}
The analysis carried on so far assumed a dilute regime, which is very different
from the conditions typical of grease ice. Furthermore, the analysis
assumed creeping flow conditions, which may be problematic for millimiter size
particles. As regards the first issue, the differential scheme, 
originally proposed by (\cite[Bruggeman
1935]{bruggeman}), provides an analytical method to generalize the
well-known Einstein formula for the effective viscosity of dilute
suspensions to finite concentrations (\cite[Brinkman
1952]{brinkman}), (\cite[Roscoe 1952]{roscoe}). More recently,
the differential scheme has been exploited to study the viscosity
problems related to randomly oriented spheroidal inclusions in
viscous fluids (\cite[Phan-Thien \& Pham 2000]{pt}). 

The basic idea is to calculate the increment of effective viscosity
that is obtained adding a volume $\Delta v$ of particles to a volume
$v_0$ of suspension already containing a fraction $v/v_0$ of particles.
Let us indicate with $\nu=\mu/\rho_w$ the kinematic viscosity of the
solvent and with $\bar\nu\simeq\bar\mu/\rho_w$ the same quantity 
referred to the suspension, with $\rho_w$ the density of the solvent, 
assumed approximately equal to that of the suspension.
The increment in volume fraction is
$$
\Delta\phi=\frac{v+\Delta v}{v_0+\Delta v}-\frac{v}{v_0}
\simeq (1-v/v_0)\Delta v/v_0=(1-\phi)\Delta v/v_0
$$
while the increment in effective viscosity $\bar\nu$ will be
$$
\Delta\bar\nu=\frac{K\bar\nu\Delta v}{v_0+\Delta v}
$$
leading to the differential equation
\beq
\d\bar\nu/\d\phi=K\bar\nu/(1-\phi),
\qquad
\bar\nu(0)=\nu
\label{cauchy}
\eeq
The differential
scheme assumes implicitly that viscosity renormalization is the
only effect of the particle inclusion, which is strictly true only
when the particle orientation distribution and consequently the
viscosity tensor are isotropic. In the opposite limit $\homega\le 1$,
the theory of (Sundararajakumar \etal1994) could be applied.

In the case of microscopic (Stokesian) particles,
Eq. (\ref{cauchy}) could be integrated from the initial condition (corresponding
to pure solvent) to the final concentration $\phi$, using for $K$ the
constant value obtained from
Eqs. (\ref{ABC},\ref{Knorot},\ref{Krot}) and the data in Fig.
\ref{icefig2}. In this case, the solution would be
\beq
\bar\nu(\phi)=\nu(1-\phi)^{-K},
\label{diff_scheme}
\eeq
(This equation allows a maximum packing fraction $\phi_{\rm MAX}=1$ which is above
the real value $\phi_{MAX}\simeq\pi/4$ appropriate for stacked discs).
In the case of frazil particles, in the integration of Eq. (\ref{cauchy}),
there is an initial range of values of $\phi$ in which $\bar\nu(\phi)$ is 
likely to be too small for the creeping flow approximation to apply; in 
that range, the stress coefficients and $K$ will likely differ from their
zero Reynolds number limit. Hence $K=K(\phi)$ and Eq. (\ref{diff_scheme}) 
will provide at most an order of magnitude estimate for the effective 
viscosity of grease ice.

To determine the importance of these effects,
let us consider an individual ice platelet in a suspension of effective
viscosity $\bar\nu$. Creeping flow conditions require stationarity and 
small particle Reynolds numbers. Introducing the effective Stokes time 
$\bar\tau_S\sim b^2/\bar\nu$:
\beq
\omega\bar\tau_S=\frac{\omega b^2}{\bar\nu}\ll 1,
\qquad
\Rey_p=\frac{eb^2}{\bar\nu}=\omega\bar\tau_Sk\A\ll 1,
\label{conditions}
\eeq
where $e=k\widetilde{U}$ is the strain strength.
In the case of a dilute suspension, $\bar\nu=\nu\simeq 0.01\cm^2\sec^{-1}$;
taking for the particle radius the value $b\sim 0.1\cm$,
we would obtain from Eqs. (\ref{dispersion},\ref{utilde}), 
values for $\omega\bar\tau_S$ ranging from
$1$ in open sea, to $10$ in laboratory conditions. As regards the condition on $\Rey_p$,
this is satisfied in open sea, where $k\A$ is of the order of a few hundredths, but
only marginally in the laboratory, where $k\A\sim 0.3$ (Martin \& Kauffman 1981). 
Clearly, the conditions of Eq. (\ref{conditions}) are going to be satisfied in the
case the suspending medium is the grease ice, when $\bar\mu\sim 10^2\cm^2\sec^{-1}$.
This has the consequence, in particular, that the results on the
orientation dynamics of Section II are expected to remain overall valid.

In the dilute case, the non satisfaction of the conditions in
Eq. (\ref{conditions}) has another consequence, namely,
inertia will cause relative particle-fluid motions and additional dissipation 
in the suspension. We can obtain an estimate of this effect. The particle
velocity $\V$, subtracted of the contribution from the buoyancy produced 
drift, will obey an equation in the form:
\beq
\Big(\frac{\d}{\d t}+\frac{1}{\bar\tau_S}{\bf\Pi}\cdot\Big)(\V-\U)\simeq
\epsilon r\frac{\d\V}{\d t}
+\frac{1}{\bar\tau_S}{\bf F}\cdot \U+{\bf f}
\label{maxey}
\eeq
with $\epsilon=1-\rho_p/\rho_w\simeq 0.1$, $\rho_p$ indicating the ice
density, ${\bf F}=O((kb)^2)$ accounting for the Faxen 
force and
${\bf\Pi}$ the adimensionalized resistance tensor, whose components are $O(1)$ for the
range of $\omega\bar\tau_S$ we are interested in (Zhang \& Stone 1998). The term ${\bf f}$ is
a noise contribution accounting for collision effects with other particles and Brownian
motion. All inertia effects in the wave flow and from the particle relative motion are 
in first instance neglected.

The contribution to relative motion from collisions is important in bubbly flows 
(\cite[Kang et Al. 1997]{kang}), but can be shown to be negligible in the 
present case due to the regime $\Rey_p<1$. Let us show this. 
We can estimate the noise amplitude
in a kinetic approach by introducing first the collision frequency $\tau_c^{-1}$
$$
\tau^{-1}_c\sim n\Delta V b^2=\frac{\phi\Delta V}{a}
$$
where $n=\phi/ab^2$ is the numerical density of the particles, $\Delta V\sim |\V-\U|$
estimates the typical collision velocity and $b^2$ estimates the collision cross section.
Collisions will be important provided $\tau_c<\bar\tau_S$; in this case, taking the noise
as uncorrelated $\langle f(t)f(t')\rangle\sim D\delta(t-t')$, we would have for 
its amplitude:
$$
D\sim\Delta V^2\tau_c\sim\frac{a\Delta V}{\phi}
$$
From Eq. (\ref{maxey}), considering $f$ dominant on the other terms to right hand side, 
we obtain the estimate for the velocity 
$\Delta V\sim e(D\bar\tau_S)^\frac{1}{2}$; this is the 
velocity difference in the wave field sampled by the diffusing particle
in the relaxation time $\bar\tau_S$. Substituting the expressions for
$D$ and $\bar\tau_S$, we find 
$$
\Delta V\sim \frac{ab^2e^2}{\phi\bar\nu}
$$
and substituting in the expression for $\tau_c$ and comparing with $\bar\tau_S$, we
see that the condition $\bar\tau_S>\tau_c$ is equivalent to $\Rey_p>1$. 
Interparticle collisions can then be neglected.

Passing to the contribution to $\V-\U$ from direct acceleration by the wave field,
we see that, for waves in both laboratory ($k\sim 10^{-1}\cm ^{-1}$) 
and open sea ($k\sim 10^{-3}\cm^{-1}$) conditions, the Faxen force can be disregarded.
Neglecting the noise in Eq. (\ref{maxey}), we obtain in this limit:
$$
|\V-\U|\sim\epsilon r\widetilde{U}\, {\rm min}(1,\omega\bar\tau_S)
$$
The dissipation produced by a single particle due to its translation relative to the fluid can
be estimated from the product of the drag force in Eq. (\ref{maxey}) and $|\V-\U|$ 
as $\rho_wb^3\Pi|\V-\U|^2/\bar\tau_S$. The contribution from relative particle fluid 
rotation will be smaller by a factor $kb$. The dissipation per unit volume will be therefore:
$$
W_{tr}\sim \bar\mu\phi r\Big(\frac{\epsilon\widetilde{U}}{b}\Big)^2 
{\rm min}(1,\omega\bar\tau_S)^2
$$
to be compared with the viscous dissipation $W_\nu=\bar\mu(k\widetilde{U})^2$.
We thus see that the contribution to dissipation from relative particle-fluid motion
is dominant in the dilute case, but decreases with $1/\bar\mu$ when $\omega\bar\tau_S<1$.
Taking for $\bar\mu$ values in the range of the hundreds, and $\phi\sim 0.3$, as 
observed in grease ice (see also Table 1), we see that the contribution to
dissipation from relative particle-fluid motion is negligible, with the ratio 
$W_{tr}/W_\nu$ varying from $10^{-4}$ in the open sea to $10^{-8}$ in the
laboratory. 

In conclusion, inertia and non-stationarity are likely to contribute to the 
effective value of the grease ice effective viscosity, but do not affect 
the frazil orientation dynamics described in Section II. It is also 
confirmed that the 
dominant contribution to dissipation and to the effective viscosity is 
the particle induced stress and not particle-fluid motions.

\section{Comparison with experiments}
We have compared the order of magnitude estimate for the
effective viscosity, provided by Eq. (\ref{diff_scheme}), with
the wave tank data of (\cite[Martin \& Kauffman 1981]{mk}). In
their experiment, concentrated suspensions of grease ice with
thicknesses varying from $7$ to $15\cm$ and volume fraction $\phi$ between
$.28$ to $.44$, were allowed to grow in a
$2$m long tank previously filled with saline water to a depth
of $41\cm$. 
%The waves were generated by means of a paddle, mounted
%at one end of the tank.
We selected those measurements relevant to
propagation of deep waves in grease ice layer according to the
criterion $kh\geq\pi/2$, where $k$ is the open water wavenumber
and $h$ is the ice layer thickness (\cite[Phillips
1966]{phillips}). 
The relevant parameters of the experimental data we are considering
are listed in Table 1.
As already discussed, the differential
scheme assumes an isotropic suspension. Comparing the values of
$\hat\omega$ in Table 1 with Figure \ref{icefig3}, 
we see that the data fall in the range where this
assumption holds. 

\begin{tabular}{cccccccc}
\hline

$\hat\omega$& $\phi$ & $\omega (\sec^{-1})$ & ${\cal A} (\cm)$ &  $h (\cm)$ 
& $q ({\rm m}^{-1})$ & $\bar{\nu} (\cm\times\sec^{-2})$ & $K$ \\

$1.4\div 1.5$ 
& $0.30 \div 0.34$ & $14.9$ & $1.45 \div 1.55$ & $6 \div 7$ & $5.3\pm0.4$ 
& $243 \div 250$ 
& $18.7\pm0.2$ \\

$1.3\div 1.4$ 
& $0.29 \div 0.34$ & $14.9$ & $1.6 \div 1.7$ & $7 \div 8$ & $6.6\pm0.6$ 
& $353 \div 361$ 
& $19.7\pm0.3$ \\

$1.2\div 1.3$ 
& $0.28 \div 0.32$ & $14.9$ & $1.7 \div 1.8$ & $8 \div 9$ & $5.8\pm0.5$ 
& $275 \div 280$ 
& $20.3\pm0.2$ \\

$1.2\div 1.5$ 
& $0.28 \div 0.34$ & $14.9$ & $1.5 \div 1.8$ & $7 \div 8.5$ & $1.6\pm0.1$ 
& $587 \div 603$ 
& $15.6\pm0.1$ \\

$1.2\div 1.5$ 
& $0.34 \div 0.35$ & $14.9$ & $1.5 \div 1.8$ & $8.5 \div 10$ & $7.5\pm0.5$ 
& $454 \div 463$ 
& $18.6\pm0.3$ \\

$1.4\div 1.4$ 
& $0.35 \div 0.44$ & $10.7$ & $3.0 \div 3.1$ & $14 \div 16$ & $3.5\pm0.2$ 
& $1010 \div 1029$ 
& $17.2\pm0.2$ \\

\multicolumn{8}{c}{Table 1}\\
\hline
\end{tabular}

\noindent
Grease ice effective viscosities were estimated from the measured spatial
attenuation rate
$q=-{\cal A}^{-1}\d{\cal A}/{\d x_1}$ 
using a two-layer viscous fluid wave
propagation model (\cite[De Carolis \& Desiderio 2002]{dec}). 
The model can be inverted to 
obtain the effective viscosity $\bar\nu$ of the grease ice from
the experimentally observed values of the wave frequency $\omega$, the 
attenuation rate $q$ and the thickness $h$ and volume 
fraction $\phi$ of the frazil. 
The reduced viscosity
$K$ can then be determined from Eq. (\ref{diff_scheme}) in
order to estimate the corresponding particle aspect ratio by
means of Eqs. (\ref{ABC},\ref{Knorot},\ref{Krot}) and the data in Fig.
\ref{icefig2}. From the data in Table 1, we obtain the
estimate: $r\sim 2\times10^{-2}$, to be compared with the individual
observation presented in 
\cite[Martin \& Kauffman (1981)]{mk} 
of a disk diameter of $0.1\cm$  and thickness of $1-10\micron$
(see their Fig. 11). The corresponding range of
variability for frazil ice in  geophysical environment is
$0.1-0.4\cm$ in diameter and 1-100$\micron$ in thickness
(\cite[Kivisild 1970]{kivisild}), corresponding to
0.25$\times10^{-4}<r<0.1$.

\section{Conclusions}
We have obtained predictions on the effective viscosity dependence
of a suspension of disk-like particles, in the velocity field of deep
water waves, on the aspect ratio and the 
concentration of the particles. A key parameter appears to be, in the dilute
limit, the relative strength of the wave field strain, which
parameterizes the wave amplitude, and the wave frequency. For high
amplitude waves, collective alignment of the particles in
suspension with the wave field is possible, with the crossover to
this regime signalled by a deep minimum in the effective
viscosity. This minimum is actually lower than the effective
viscosity in the case of spherical particles and can be smaller,
by orders of magnitude, than the value of the effective viscosity
of a disk-like particle suspension away from the crossover. 

An interesting question is whether these behaviors are preserved away
from the deep water wave regime we have considered in this paper.
For shallow water waves, a rotating system in which the flow
becomes time-independent does not exist anymore, and irregular
behaviors of the kind described in (Szeri \etal1992) become
possible. 

Some of these results extend away from the dilute limit
to the case of grease ice, in particular,
the presence of a crossover to coherent alignment of the particles
for large amplitude waves (\cite[Martin \& Kauffman 1981]{mk}).
As regards the effective viscosity of the grease ice, this appears
to be dominated by the stress contribution from the individual 
particles rather than from momentum transport from relative 
particle-fluid motion. This is in contrast with other situations,
e.g. bubble laden flows (\cite[Kang et Al. 1997]{kang}), in 
which the second is the dominant effect. A creeping flow based
calculation of the effective viscosity, aided by the use of 
a differential scheme, to deal with the high concentration 
regime, leads to results consistent with experiments.

\vskip 10pt
\noindent{\bf Acknowledgments:}
This work was supported by the Commission of the
European Communities under contract EVK2-2000-00544 of the
Environment and Climate Programme. The authors would like to
thank Prof. G. Spiga for valuable comments and stimulating
discussions.

\end{document}